\documentclass[aps,prd,twocolumn,groupedaddress,showpacs,showkeys,nofootinbib,amssymb]{revtex4}
\usepackage[latin1]{inputenc}
\usepackage{graphicx}
\usepackage[english]{babel}
\usepackage{amsmath}
\usepackage{amssymb}
\usepackage{amsfonts}
\usepackage{colordvi}
\usepackage{color}

\begin{document}

\title{MOND's acceleration scale as a fundamental quantity}

\author{Tula Bernal$^{1}$, Salvatore Capozziello$^{2,3}$,  Gerardo Cristofano$^{2,3}$, Mariafelicia De Laurentis$^{2,3}$}

\affiliation{\it $^1$Instituto de Astronom\'{\i}a, Universidad Nacional
           Aut\'onoma de M\'exico, AP 70-264, Distrito Federal 04510,
           M\'exico,
           $^2$Dipartimento di Scienze Fisiche, Università di Napoli ``Federico II'',
           Napoli, Italy,
           $^3$INFN Sez. di Napoli, Compl. Univ. di Monte S. Angelo, Edificio G,
           Via Cinthia, I-80126, Napoli, Italy}

\date{\today}

\begin{abstract}
Some quantum-cosmic scaling relations indicate that the MOND acceleration parameter $a_0$ could be a fundamental quantity ruling the self-gravitating structures, ranging from stars and globular clusters up to superclusters of galaxies and the whole observed universe. We discuss such coincidence relations starting from the Dirac quantization  condition ruling the masses of primordial black holes.

\end{abstract}

\keywords{Alternative theories of gravity; cosmology; large scale structure; quantization conditions.}
\pacs{04.50.+h, 95.36.+x, 98.80.-k}

\maketitle

\section{Introduction}
\label{introduction}

 The  observed scales of astrophysical self-gravitating
systems, like galaxies, galaxy clusters  and the
universe itself, constitute a puzzle of  modern
physics. From a fundamental  point of
view, it is expected that all bounded  self-gravitating  structures  have a common  origin related to  the
 cosmic microwave background radiation.  Such structures should bring  the signature of  primordial
quantum fluctuations   which should have determined
the scales of systems that  we observe  today.

 The main goal of  studies dealing with this issue is to frame the large scale structure into some
unifying theory in which all the today observed structures can be treated
under the same fundamental standard. In this sense, the remnants of the
primordial epochs, probed by the cosmological observations, are the ideal test
 to constrain such  theories.

Looking for a fundamental theory connecting the microscopic to
the macroscopic scales, many ``coincidences" have been identified among cosmic
and fundamental parameters, like the cosmological constant, the gravitational
constant, the speed of light and the Planck constant. The  so-called Large Numbers
Coincidence (LNC) \cite{lnc}, for example, refers to  some
fundamental relations  to pure numbers of order
$10^{40}$. Eddington and Dirac hypothesized that such coincidences could be
dynamically generated by the cosmological parameters related to quantum ones. Furthermore, this coincidence seems to 
to occur at the same epoch in which other coincidences related to
cosmic parameters occur. This fact   suggests some   underlying physical law that should be fully understood
\cite{funkhouser06}.

 In addition, there is the issue involving the so-called ``dark"
components, i.e.  dark matter and dark energy, being approximately the $95\%$
of the total mass-energy content of the universe. In principle, the dark
matter was postulated to explain the observed rotation curves of  spiral galaxies  \cite{zwicky37}; then it was necessary to explain
the mass to light ratios of  galaxies and galaxy clusters, the
gravitational lensing,  the structure formation and most of  astrophysical phenomena and structures that escape the standard description by Einstein (Newton) gravity 
and luminous matter \cite{binney08}, 
 On the other hand,  the dark
energy has been postulated in order to explain the today observed accelerated expansion rate of the
Hubble cosmic fluid, firstly deduced by the supernovae Ia observations \cite{SNIa1, SNIa2}.

As alternatives to the dark matter and energy problems, several theories
modifying the gravity, instead of  the inclusion of unknown 
components, have been proposed, in both non-relativistic and relativistic
regimes, indicating a failure in the Newtonian and general relativistic
theories of gravity  at large scales (see e.g.
\cite{book, mendoza10}).

 In the non-relativistic limit, Milgrom introduced the MOdified Newtonian
Dynamics (MOND) \cite{milgrom83,milgrom08,milgrom09} to explain the observed rotation
curves of spiral galaxies, through the inclusion of a phenomenological
acceleration scale $a_0$ such that, for accelerations much greater than $a_0$,
the dynamics is Newtonian and for accelerations much less than $a_0$, the
dynamics is modified and reduces to constant circular velocities at large radii.

The MOND frame has been proved to be successful on many astrophysical
situations, but difficult on others (see e.g. \cite{mendoza10} and references therein).
Despite these facts, the key feature of this modification, is the
introduction of an acceleration scale $a_0$ into the gravitational interaction \cite{binney08}.
Concerning  $a_0$,  some of the mentioned
coincidences can be considered involving the Hubble constant,  the  cosmological constants, and the
 the nucleon size \cite{funkhouser06}.
Quantum mechanics and quantum field theory  play a main role in 
understanding such coincidences showing that $a_0$ could be actually considered 
among the number of fundamental parameters of physics.

In  this paper, we will face this problem. Starting from a quantum field theory approach, a quantum relation for primordial
black holes has been proposed  whose validity extends
from  the Planck scales to the today observed astrophysical structures \cite{capozziello10a}. 
It can be proven that such a  quantum relation is  a 
scaling one  containing, in principle,  the signature of  astrophysical
structures starting from  their basic constituents, the nucleons \cite{capozziello10b}.
Specifically, it  reproduces
 the so-called  Eddington-Weinberg relation \cite{eddington-weinberg}, as well as
 the phenomenological statistical hypothesis for  self-gravitating
systems \cite{scott1,scott2}, where the characteristic
sizes of astrophysical structures can be recovered assuming that gravity is
the overall interaction assembling systems from  fundamental microscopic
constituents.

 In this paper, we  show how, starting from  fundamental scaling relations
involving cosmological coincidences,
MOND's acceleration emerges as a   characteristic
scale for  self-gravitating astrophysical systems. We take into account only  characteristic  masses,  radii and
the baryonic components. It is worth noticing  that the acceleration  scale is addressed without assuming any  dark component.

 The  paper is organized as follows. In Sec. \ref{scaling-relations}, we
recall the  scaling hypothesis working, in principle,  for all  self-gravitating
structures, from  Planck's scales to the universe itself. From the scaling relations, a characteristic
acceleration is deduced  and  identified as $a_0$ in terms of  fundamental quantities.

 In Sec. \ref{signature}, the signature of  scaling relations for several
self-gravitating systems is discussed  as a consistency check to obtain the
characteristic sizes of the structures from their  characteristic baryonic masses.

  Sec. \ref{mond}  is  devoted to a brief  summary on
how the acceleration scale $a_0$  modifies the Newtonian dynamics. In Sec. \ref{coincidences},
we  discuss the  coincidence relations  showing that   $a_0$ can be considered  among  the
fundamental parameters of scaling relations.

 Discussion  and conclusions are reported in  Sec. \ref{discussion}.

\section{Scaling hypothesis for astrophysical systems}
\label{scaling-relations}

Assuming the  Dirac quantization condition \cite{dirac31} and a quantum  field 
theory result relating the electric and magnetic charges of primordial 
black holes to their masses $M$, the following quantization relation can be obtained \cite{capozziello10a}:

\begin{equation}
  G M^2 = n \hbar c \, ,
\label{qbh}
\end{equation}
where $G$ is the gravitational constant, $n$ is the quantization
number and $\hbar$ is the reduced Planck constant. For $n=1$, the lowest mass
allowed for a quantum black hole is the Planck mass.

Moreover, it can be  shown that the above relation is valid for self-gravitating
 astrophysical structure, from globular clusters to the observed universe
itself \cite{capozziello10a}, and numerical agreement has been found with
the statistical hypothesis (based on  phenomenological considerations) applied to
 self-gravitating systems, defined as  bound states with a very large
number $N_{\textrm{as}}$ of constituents \cite{scott1,scott2}.
This connection can be equivalently obtained either by considering the protons
as the elementary constituents or, as usual in astrophysics, by
considering stars as the granular components of galaxies, and the galaxies the granular 
components of galaxy clusters and superclusters  (``granular approximation").

 In this context, the working hypothesis is that the total action of the astrophysical 
self-gravitating system, $A_{\textrm{as}}$, can be achieved from the
Planck constant as 

\begin{equation}
  A_{\textrm{as}} \approx \hbar N_{\textrm{as}}^{3/2} \, .
\label{action}
\end{equation}
This means that  the total action is given by $A_{\textrm{as}} = n_{\textrm{as}} \hbar$,
where $n_{\textrm{as}}$ is the value of the quantum level number appearing in
Eq. \eqref{qbh}. By comparing this number with the corresponding value
$N_{\textrm{as}}^{3/2}$ from Eq. \eqref{action}, a numerical scaling 
with characteristic sizes   of globular clusters,  galaxies, galaxy clusters  and the
observed universe can be achieved  \cite{demartino}.

However, it is interesting to note  that the quantization relation \eqref{qbh}  scales "analytically" for any astrophysical system down to the Planck
mass. In fact, the quantum relation can be rewritten in a
more suitable form as \cite{capozziello10b}
 
\begin{equation}
  G M_{\textrm{as}}^2 = \left( \frac{ N_{\textrm{as}} }{ N_{\textrm{BH}} } \right)^2
     \hbar c \, ,
\label{}
\end{equation}

\noindent where $N_{\textrm{BH}}$ is the number of protons in primordial
black holes.

Writing the mass of the self-gravitating system considering  its number of protons as

\begin{equation}
  M_{\textrm{as}} \approx N_{\textrm{as}} \, m_p \, ,
\label{mass}
\end{equation}

\noindent one gets simply back $G M_{\textrm{Pl}}^2 = \hbar c$, and the scaling
hypothesis \eqref{action} is immediately verified \cite{capozziello10a}.

 Another interesting outcome obtained by  quantization relation
\eqref{qbh}, can be achieved   by using the relation

\begin{equation}
  \frac { G M_{\textrm{as}} } { R_{ \textrm{as} }^2 } 
    \equiv 2 \pi a_* \, ,
\label{scaling}
\end{equation}

\noindent where $R_{\textrm{as}}$ is the characteristic radius of the
astrophysical structure and the  acceleration $a_*$ can be shown to be a universal constant.

In fact, it is possible to derive the size of the astrophysical structure
in terms of the number of its fundamental  constituents (protons), $N_{\textrm{as}}$,
and the fundamental scale of the proton, the Compton wavelength $\lambda_p$. It is

\begin{equation}
  R_{\textrm{as}} \approx 10 \sqrt{ N_{\textrm{as}} } \, \lambda_p \, .
\label{statcorr}
\end{equation}

 Such a result reproduces the statistical hypothesis relation proposed in
Refs. \cite{scott1,scott2}  with a correcting factor of
order 10 coming from statistical uncertainties \cite{capozziello10b}.

 By using the relation \eqref{mass} for the mass $M_{\textrm{as}}$  and the size $R_{\textrm{as}}$ from Eq.
\eqref{statcorr}, we can rewrite the  scaling relation
\eqref{scaling} as

\begin{equation}
  \frac { G M_{\textrm{as}} } { R_{ \textrm{as} }^2 } =
    \frac { G \, m_p } { ( 10 \lambda_p ) ^2 } \equiv 2 \pi a_* \, .
\label{scal-prot}
\end{equation}

 Also, by considering the case of the observed universe and using the Raychaudhuri
equation \cite{ray}, the last relation can be written as

\begin{equation}
  \frac {G M_{\textrm u}} {R_{\textrm u}^2} = \frac{c^2} {R_{\textrm u}}
     = 2 \pi a_* \, .
\label{scal-univ}
\end{equation}

 It is interesting to notice that specializing Eq. \eqref{statcorr} to the
radius of the universe and using Eqs. \eqref{mass} and \eqref{scal-univ}, the
Eddington-Weinberg relation for the radius of the universe is easily obtained
with the correcting factor 10:

\begin{equation}
  h = \frac{1}{10} \sqrt{ G R_{ \textrm{u}}  \, m_p^{3} }\, .
\label{e-w}
\end{equation}

 In addition, from Eq. \eqref{scal-prot},
the value of the acceleration $2 \pi a_*$ is found to be 

\begin{equation}
  2 \pi a_* = \frac {G \, m_p^3 \, c^2} {(10 h)^2} \, ,
\label{a0}
\end{equation}
which is a constant. Evaluating  $a_*$  in the last equation, we find the value
$a_* \approx 1.01 \times 10^{-10} \, \textrm{m/s}^2$.

It is important to stress that the equation for the constant acceleration  is valid
for any  astrophysical system where  Eq.\eqref{statcorr}
gives its characteristic radius and Eq.\eqref{mass} gives its mean mass.

 As we will see, the derived value for $a_*$ is very close to the
phenomenological value obtained for the constant acceleration scale
introduced in the MOND theory, that is $a_* \approx a_0$. However the value of $a_*$
 has been obtained here combining the fundamental
quantities $G$, $h$, $c$ and $m_p$. The meaning  of this result
is that the acceleration  $a_0$ could be  a fundamental quantity as we want to show in the discussion below.

\section{ Astrophysical systems from scaling relations}
\label{signature}

 A self-gravitating astrophysical system is a gravitationally bounded
assembly of stars and gas, and other point masses that can be neglected with
respect to the main stellar and gas content. These systems vary over more
than fourteen orders of magnitude in size and mass, from individual stars to
star clusters containing $10^2$ to $10^6$ stars, through galaxies containing
$10^5$ to $10^{12}$ stars, to vast clusters containing thousands of galaxies
\cite{binney08}.

The properties of these self-gravitating structures can be deduced by assuming
them to be relaxed and virialized systems where gravity is the only overall
interaction \cite{binney08}. First, we define globular clusters, galaxies and
clusters of galaxies, considering their typical lengths and masses. For the
characteristic mass we will only take the baryonic one and, as the boundary
(``size")  of these systems is not univocally defined, we will suppose a
characteristic length for the baryonic mass gravitationally bounded to the
system. It is important to remark that we are not assuming a dark matter
component for the systems. For example, for a spiral galaxy, we take a typical
radius of $\sim$10 kpc, which is the typical radius for the visible disk,
not $\sim$100 kpc for the hypothetical dark matter halo.

 Now, a star is not a purely self-gravitating system since, inside it, gravity
is balanced by the pressure due to electromagnetic and nuclear interactions.
However, we can find a characteristic gravitational length for the gravitating
objects around a star (as the size of a ``planetary system" around it). Of
course, this depends on the environment in which the star is embedded, but we
can assume a typical one, as the Solar neighborhood \cite{binney08}.

 To obtain the characteristic interaction lengths for the systems, we are using
the granular approximation, considering the stars as the granular constituents
of the astrophysical structures, whose main constituents are the protons.

 The globular clusters are very massive stellar systems, containing up to
$10^6$ stars ($M_{\textrm{gc}} \sim 10^6 \, \textrm{M}_\odot$). The typical
radii are of the order $\sim$10 pc. These systems are assumed completely
virialized due to collisional interactions between stars \cite{binney08}.

 A galaxy is a collisionless gravitating system, with masses ranging from
$M_{\textrm{gal}} \sim 10^{8} \div 10^{9} \, \textrm{M}_\odot$, for dwarf galaxies,
and $M_{\textrm{gal}} \sim 10^{10} \div 10^{12} \, \textrm{M}_\odot$, for giant
galaxies. The sizes are not well defined since there is no effective
boundary for the systems. Astronomers define operative characteristic sizes as
the \it{effective radius}, \rm $R_e$, containing half of the total luminosity
or the \it{tidal radius}, \rm $R_t$, defined as the distance from the center
where the density drops to zero \cite{binney08}. Assuming as a typical
interaction length $R_{\textrm{gal}} \sim 1 \div 10$ kpc is quite reasonable for systems ranging from
dwarf to giant galaxies.

 Groups of galaxies are systems containing $10 \div 20$ galaxies, but they are
not considered self-gravitating systems because they are always part of more
extended self-gravitating structures (clusters of galaxies). For example, our
Local Group is a part of the Virgo Cluster.

 The clusters of galaxies are the largest self-gravitating structures in the
universe. The biggest clusters have masses $\sim 10^{15} \, \textrm{M}_\odot$
within $\sim$1 Mpc from their centers.

 The superclusters of galaxies are not considered effective self-gravitating systems
because of their large sizes. For this reason, they are considered as expanding with the
expansion rate of the universe. It is supposed that all the clusters of
galaxies are part of a larger supercluster.

 Finally, the mean number of atoms (baryons) in the \it{observable universe} \rm is
supposed to be $\sim 10^{80}$, corresponding to a mass of the order
$M_{\textrm{u}} \sim 10^{23} \, \textrm{M}_\odot$. Furthermore, the estimated size
for the observed universe vary between $R_{\textrm{u}} \sim 10^{25} \div 10^{27}$ m.

 Let us now show how  the scaling relation
\eqref{statcorr} holds for astrophysical structures. To this purpose, we write the mean radius
$R_{\textrm{as}}$,  coming from Eq. \eqref{statcorr}, in terms of the mean
baryonic mass $M_{\textrm{as}}$, given by Eq. \eqref{mass} for the given astrophysical structure, as

\begin{equation}
  R_{\textrm{as}} = 10 \left( \frac{M_{\textrm{as}}}{m_p^3} \right)^{1/2}
    \frac{h}{c} \, .
\label{r-m}
\end{equation}
From this relation, we obtain, for a globular cluster with mean mass
$M_{\textrm{gc}} \sim 10^6 \textrm{M}_\odot$, the characteristic radius
$R_{\textrm{gc}} \sim 15 \textrm{pc}$.

 For a typical giant galaxy with mean mass
$M_{\textrm{gal}} \sim 10^{11} \, \textrm{M}_\odot$, the corresponding
characteristic radius is $R_{\textrm{gal}} \sim 5$ kpc.

 For a cluster of galaxies with mean mass
$M_{\textrm{cg}} \sim 5 \times 10^{14} \, \textrm{M}_\odot$, we have a
characteristic radius of $R_{\textrm{cg}} \sim 0.5$ Mpc.

 For the universe, assuming the mean mass
$M_{\textrm{u}} \sim 10^{23} \, \textrm{M}_\odot$, we obtain the mean
radius $R_{\textrm{u}} \sim 10^{26}$ m.

 In this sense, we can calculate the corresponding characteristic interaction
radius of a star with mean mass $M_{\textrm{star}} \sim 1 \, \textrm{M}_\odot$,
as $R_{\textrm{star}} \sim 3 \times 10^3 \textrm{au}$, where "au" is the astronomic unit.. This radius 
corresponds to the radius of the  {\it Oort cloud},
 which defines the typical outest gravitational boundary of our Solar System. 

 For the proton, we have that its characteristic interaction radius, from the scaling
relation \eqref{statcorr}, is given by $R_{\textrm{p}} \approx 10 \lambda_p$,
and its characteristic acceleration scales as  Eq. \eqref{scal-prot}.

\section{The MOND acceleration parameter}
\label{mond}

 The acceleration constant $a_0$ has been phenomenologically introduced by Milgrom
\cite{milgrom83} as a cut-off parameter in the MOND theory to
discriminate between Newtonian gravity and modified dynamics. MOND is constructed to obviate the need of dark matter when applied to
galactic systems, for which the standard Newtonian dynamics is a good
approximation only for accelerations much larger than $a_0$, and the so-called
deep MOND regime is valid for accelerations much less than $a_0$.

 The best value for $a_0$ is obtained from the fit to the rotation curves
of spiral galaxies in the deep MOND regime, in the vicinity of our
Galaxy and for a value for the Hubble constant $H_0 = 75 \, \textrm{km} \,
\textrm{s}^{-1} \, \textrm{Mpc}^{-1}$, as $a_0 \approx 1.2 \times
10^{-10} \textrm{m/s}^2$ \cite{begeman91}. This value is taken as a constant of
the theory for all the applications to the different systems in the universe.

 At this stage, MOND is purely phenomenological, and people has done many efforts
to construct a fundamental theory from which MOND would be the correct limit for
the accelerations $a \ll a_0$ (see e.g. \cite{mendoza10} and references therein).

 In the search of such fundamental theory, Milgrom first noticed the following
coincidence between the value of the acceleration scale $a_0$, the Hubble
constant at the present epoch and the speed of light \cite{milgrom83,milgrom08}:

\begin{equation}
  2 \pi a_0 \approx c H_0 \, .
\label{H0}
\end{equation}

 From the Friedmann - Lema\^{\i}tre - Robertson - Walker cosmological equations,
the last relation can be written in terms of the cosmological constant
$\Lambda$ as (see e.g. \cite{milgrom08})

\begin{equation}
 2 \pi a_0 \approx c \left( \frac{\Lambda}{3} \right) ^ {1/2} ,
\label{lambda}
\end{equation}

\noindent since $H_0 \approx (\Lambda / 3) ^{1/2}$ at the present time.

This fact can be interpreted, from one side, as the coincidence   connecting MOND
to cosmology and  dark energy, and,  on the other side, as the influence of the cosmic
large scales  to the local dynamics \cite{milgrom08,milgrom09}.

 An additional comment to this coincidence is the fact that, as the Hubble
parameter $H_0$ varies with  cosmic time, $a_0$  varies too. This is not
necessarily the case since $a_0$ could be related to $\Lambda$ with the
latter being constant. In this case, however, the problem of the tiny value of $\Lambda$ 
with respect to the large value of gravitational vacuum state still remains (the so-called Cosmological Constant Problem).
Interestingly, variations of $a_0$ could induce secular
evolution in galaxies and other galactic systems \cite{milgrom08}. Although 
the value obtained from the rotation curve of
a spiral galaxy, at the redshift $z=2.38$, is consistent with the local measured
value from the best studied rotation curves of spiral galaxies in vicinity of
the Milky Way, this possibility is not completely discarded because of the
high uncertainties in the observations  \cite{milgrom08}. 
If $a_0$ does not vary with the cosmic time, then this coincidence
just occurs at the present epoch.

\section{``Coincidences" for the value of $a_0$}
\label{coincidences}

Assuming $a_* = a_0$,    the scaling relation \eqref{scaling}  holds
 for two very different systems:  the observed universe and the nucleon.

 In the case of the universe, it is possible to construct a length scale $l_0$ from the
constants $G$, $c$ and $a_0$, such that $l_0 \equiv c^2 / a_0 \approx 10^{27}$ m,
and a mass scale $M_0 \equiv \mu_0 c^4 \approx 6 \times 10^{23} \, \textrm{M}_\odot$
(see e.g. \cite{milgrom08}). Looking for a connection with quantum theory, these
scales can be seen as the Planck length and mass constructed from the fundamental
constants $\hbar$, $G$ and $c$. In this sense, $l_0$ and $M_0$ can give
the scales where MOND effects, combined with gravity, are expected. In fact,
 the Hubble radius and the mass of the universe can be written as

\begin{equation}
  R_H \equiv \frac{c}{H_0} \approx \frac{l_0}{2 \pi} \sim 10^{26} \, \textrm{m} \, ,
\label{hubble-rad}
\end{equation}

\begin{equation}
  M_{\textrm{u}} \approx \frac{c^3}{G H_0} \approx \frac{M_0}{2 \pi} \sim 10^{23} \, M_\odot \, .
\label{hubble-mass}
\end{equation}
From these two relations, the coincidence \eqref{H0} can be written as
\cite{milgrom08}

\begin{equation}
  2 \pi a_0 \approx \frac {G M_u} {R_H^2} \, .
\label{a0-mil}
\end{equation}
We can see that this result is exactly reproduced by Eq. \eqref{scal-univ}
because, despite of the different definitions for the radius of the universe, the
corresponding value for the Hubble radius \eqref{hubble-rad} and the value we
obtained in Sec. \ref{signature}  are of the same order of
magnitude. This is  because  the same mass scale \eqref{hubble-mass} is assumed.

 Additionally, Milgrom pointed out that $a_0$ is also the gravitational
acceleration produced by a particle of mass $\sim100 \, \textrm{MeV/c}^2$
at a distance equal to its Compton wavelength and Funkhouser \citep{funkhouser06} has studied the LNC for the fundamental
quantities and the cosmological coincidence \eqref{H0}. He  found
that there exists a critical acceleration coming from these coincidences
and has solved them proposing the following scaling law for the
cosmological constant:

\begin{equation}
  \Lambda \approx \left( \frac{8 \pi G}{3 c^2} \right)
      \frac{G m_n^2}{\lambda_n^4} \, ,
\label{lambda}
\end{equation}

\noindent where $m_n$ is the nucleon mass and $\lambda_n$ is its
Compton wavelength. This equation may be interpreted as the energy
density associated with the cosmological constant, scaled to the
gravitational energy density of the nucleon mass confined to a
sphere with its Compton wavelength as the radius.

 Evaluating the r.h.s. of the last scaling relation, a discrepancy of
$\sim$4 orders of magnitude is found with respect to the value of the
cosmological constant, $\Lambda \approx 3.9 \times 10^{-36} \, \textrm{s}^{-2}$.
This discrepancy is approximately solved by replacing $\lambda_n$ by
$b \lambda_n$, where $b$ is a constant of order 10
\cite{funkhouser06}.

 Assuming the scaling relation \eqref{lambda}, and using the Eddington - Weinberg
relation, the MOND's acceleration scale is found to be the characteristic
gravitational acceleration of the nucleon mass at its Compton wavelength,
scaled by the same factor $b=10$, that is 

\begin{equation}
  a_0 \approx \frac {G m_n} {(b \lambda_n)^2} .
\label{a0-funk}
\end{equation}
This relation is equivalent to the Milgrom result for a
particle of mass $\sim 100$ MeV, except for a factor $2 \pi$, recovered assuming  Eq. \eqref{scal-prot}.

 Finally, it is possible to show how to obtain the coincidence relation \eqref{H0} from the
quantization relation \eqref{qbh}. Rewriting such a quantum relation
in terms of the angular momentum $J$ of the mass $M$, we have

\begin{equation}
 G M^2 = n \hbar c = J c \, .
\label{qbh-j}
\end{equation}

 Now, the intrinsic angular momentum for a  mass $M$ and a
radius $R$ rotating at the angular velocity $\omega$, is given by

\begin{equation}
 J \sim \omega M R^2 .
\label{j}
\end{equation}
Writing the last two relations for the mass and radius of the observed
universe, by using Eqs. \eqref{j} in \eqref{qbh-j}, and after
rearranging terms, we have 

\begin{equation}
 \frac{G M_{\textrm u}^2}{R_{\textrm u}^2} \sim \omega_0 c \sim
    \frac{c}{T_0}  \, ,
\end{equation}

\noindent where $\omega_0$ represents the `intrinsic angular velocity'
of the universe and $T_0$ its `rotation period', that can be
approximated by the age of the universe. Then, from the Hubble constant,
$H_0 \sim 1 / T_0$, and from the scaling relation \eqref{scal-univ},
where $a_* = a_0$, we recover the coincidence given by Eq. \eqref{H0}.

 An intrinsic angular momentum for the entire universe from which this
coincidence can be derived, may have some implication for the so-called
``axis of evil" \cite{axis}.

\section{Discussion and conclusions}
\label{discussion}
In this paper, we have discussed some quantum-cosmic coincidence relations involving 
the phenomenological parameter $a_0$ which sets the so-called MOND scale.

In this perspective, such an  acceleration seems
more than only a  phenomenological parameter, but a fundamental quantity related to some universal constants (Eq. \eqref{a0}) and 
coming from a quantization condition \eqref{qbh} for quantum-gravitational systems.

From the identification $a_* = a_0$, the  scaling relations \eqref{statcorr}  and  \eqref{scaling}  
hold for any self-gravitating astrophysical structure where gravity is the overall interaction that bounds the system. 
Such relations connect the microscopic constituents (protons) with the macroscopic features of the astrophysical systems (radius and mass); the acceleration $a_0$ gives the  natural cut-off where dynamics changes regime without invoking any dark matter.
Being connected also with the Hubble parameter  $H_0$, Eq.\eqref{H0}, and the cosmological constant $\Lambda$, Eq.\eqref{lambda},  $a_0$ could give rise also to  a natural explanation 
for dark energy phenomena. 
As  a final comment,  we have to say that  the relations  presented here should be seriously considered in view to explain the universe content just with observable quantities. However,  stating that $a_0$ is a fundamental parameter, we do not pretend to be conclusive  since we need a self-consistent relativistic quantum field theory where MOND is fully recovered in the weak field limit.

\section{Acknowledgements}

TB acknowledges economic support from CONACyT 207529.

\end{document}